# Surface plasmon enhanced fluorescence: self-consistent classical treatment in the quasi-static limit

Dentcho A. Genov

College of Engineering and Science, Louisiana Tech University, Ruston, Louisiana 71272, USA

**Abstract**

The problem of enhanced molecular emission in close proximity to dielectric and metallic interfaces is of great importance for many physical and biological applications. Here we present an exact treatment of the problem from the view point of classical electromagnetism. Self-consistent analytical theory of the surface fluorescence enhancement is developed for configurations consisting of an emitter in proximity to core-shell metal-dielectric nanoparticles. The dependence of the fluorescence enhancement on the excitation laser and fluorescence frequencies and distance of the emitter to the nanoparticle interface are studied. The developed theory predicts enhanced fluorescence at intermediate distances as well as emission quenching into non-radiative surface plasmon (SP) modes dominating the response for short distances. The conditions for optimal emission enhancement for two core-shell configurations are determined. The theory can be applied toward analyzes and optimization of various applications related to SP enhance fluorescence spectroscopy.

## 1. Introduction

The fluorescence emission is a common marker for a wide variety of practical problems including fluorescence tagging related to bio-sensing and spectroscopic studies [1-5], dye and quantum dot lasers [6,7], and electrochemistry [8]. When an emitter is placed in close proximity to a plasmonic interface enhanced emission is expected, a process also known as Purcell enhancement [9]. This effect is especially prominent in low-dimension metallic nanostructures [10] and has been extensively utilized to observe naturally weak physical effects with enhanced sensitivity including surface enhanced fluorescence (SEF) and Raman spectroscopy [11-14]. Furthermore, applications in laser and nanolaser systems and as ultrafast light sources have been considered [15-19].

The theoretical treatment of surface enhanced fluorescence (SEF) has a long history. Analytical models from the viewpoint of classical electromagnetic theory have been developed for semi-infinite metal interfaces [20, 21], spherical and spheroidal metal particles [22-25], dimers [26, 27] and embedded concentric spheres [28-30]. Studies based on the quantum electrodynamics approach have been also proposed and carried out in details [31-34]. Within the applicability of the linear response theory (weak coupling) the classical and quantum approach have been shown to lead to similar results, with emission rates governed by competing radiative and non-radiative processes. It must be pointed out that the vast majority of presented theories consider a problem of single dipole emitter with fixed dipole moment in the presence of metal structures. The general case in which the molecular dipole moment is self-consistently obtained through the molecular polarization and the local electromagnetic field consisting of the incident laser radiation (pump) and scattered field due to the nanostructures, is not well studied. To address this problem, in this paper we present, the first completely self-consistent model of core-shell nanoparticles in the present of point dipolar emitters from the point view of classical electromagnetism.

The organization of the paper follows. In section 2, we present the classical electromagnetic theory of fluorescence emission enhancements of a single molecule placed next to a three-dimensional core-shell particle. First, we provide an approximate result in which the nanoparticle and emitter are treated as simple electric point dipoles. The model properly predicts the fluorescence enhancement for emitters at sufficiently large distances from the nanoparticle. A complete, self-consistent multipole theory is then developed under the quasi-statics approximation. In section 3, the fluorescence emission enhancement is analyzed in term of the molecular-particle separation distance, filling fraction of the nanoparticle core-shell configuration and frequency of illumination/emission. The conclusions are finally drawn in section 4.

## 2. Theoretical description

The system under consideration is presented in figure 1 and consists of a single molecular emitter situated at a distance $R$ along the $+z$ axis from a core-shell metal-dielectric nanoparticle. The nanoparticle is positioned at the center of the coordinate system and has inner core radius $b$ and outer shell radius $a$. In the model the emitter-nanoparticle system is driven by an external arbitrary polarized time harmonic electric field $\vec{E}(t) = \vec{E}_0 e^{-i\omega t}$ and the molecular emitter is modeled as a point dipole with dipole moment $\vec{p}_d(t) = \vec{p}_d(\omega)e^{-i\omega t}$. Neglecting the Stokes shift and restricting our analysis to the linear response theory we have $\vec{p}_d(\omega) = \alpha_d(\omega)\vec{E}_d(\omega)$, where $\vec{E}_d$ is the electric field amplitude at the position of the dipole (excluding the dipole self-field). The molecular polarizability is give as $\alpha_d(\omega) = 4\pi\epsilon_0 a_d^3 \mathcal{L}_d$, where $\mathcal{L}_d = \omega_0^2/(\omega_0^2 - \omega^2 - i\omega\omega_{\tau,d})$ is the molecular line shape function, $\omega_0$ is the fluorescence frequency, $\omega_{\tau,d}$ is the damping frequency, and we have introduced an effective molecular dipole radius $a_d = (e^2/4\pi\epsilon_0 m\omega_0^2)^{1/3} = (d^2/2\pi\epsilon_0 \hbar\omega_0)^{1/3}$ related to the transition dipole moment $d = |\langle 2|e\vec{r}|1\rangle|$. Provided the distance between nanoparticle and emitter satisfies the condition $\omega R \ll c$, we can threat the problem in the quasi-static approximation involving the corresponding quasi-static electrical potentials and fields. In what follows, the SEF response is first considered in the effective medium dipole-dipole approximation and then a closed form, self-consistent multipole theory is presented.

### 2.1. Coupled dipole-dipole approximation

If the distance between emitter and nanoparticle is sufficiently large $R \gg a$, we can consider the problem as a simple dipole-dipole system coupled to the incident laser radiation. Under this condition the nanoparticle is modeled as a point dipole situated at the center of the coordinate system (see figure 1(b)) having a dipole moment $\vec{p}_p(\omega) = \alpha_p(\omega)\vec{E}_p(\omega)$, with $\vec{E}_p$ being the electric field amplitude at the nanoparticle position (excluding the particle self-field). The particle polarizability is $\alpha_p(\omega) = 4\pi\epsilon_0 a^3 \mathcal{L}_p$, where $\mathcal{L}_p = (\epsilon_e - \epsilon_0)/(\epsilon_e + 2\epsilon_0)$ is the dipolar polarizability line shape function represented through the dipolar effective permittivity of the core-shell nanostructure [35,36]

$$\epsilon_e = \epsilon_2 \frac{2\epsilon_2 + \epsilon_1 - 2p(\epsilon_2 - \epsilon_1)}{2\epsilon_2 + \epsilon_1 + p(\epsilon_2 - \epsilon_1)}, \quad (1)$$

where $p = (b/a)^3$ is the core volume filling fraction. The external electric fields at the positions of the particle and molecular emitter are then given by the coupled equations

$$\vec{E}_d = \vec{E}_0 + \frac{\alpha_p}{4\pi\epsilon_0 R^3}\left(3(\hat{z}\cdot\vec{E}_p)\hat{z} - \vec{E}_p\right)$$
$$\vec{E}_p = \vec{E}_0 + \frac{\alpha_d}{4\pi\epsilon_0 R^3}\left(3(\hat{z}\cdot\vec{E}_d)\hat{z} - \vec{E}_d\right) \quad (2)$$

Since the radiated power by the molecule is $P_d \sim \omega^4 |\alpha_d \vec{E}_d|^2$ the fluorescence emission enhancement is simply proportional to the enhancement of the local electric field intensity at the position of the molecule. Solving equation (2) for the fields and averaging over the incident field polarizations the SEF factor follows as

$$F = \frac{\langle |\vec{E}_d|^2 \rangle}{|\vec{E}_0|^2} = \frac{2}{3}\left|\frac{1 - \mu_p \mathcal{L}_p}{1 - \mu_p \mu_d \mathcal{L}_p \mathcal{L}_d}\right|^2 + \frac{1}{3}\left|\frac{1 + 2\mu_p \mathcal{L}_p}{1 - 4\mu_p \mu_d \mathcal{L}_p \mathcal{L}_d}\right|^2, \quad (3)$$

where $\mu_p = (a/R)^3$ and $\mu_d = (a_d/R)^3$ are geometrical factors. The first and second terms in equation (3) correspond to contributions due to molecular polarization parallel and normal to the particle surface, respectively. We must note that similar expression for parallel field polarization has been already obtained for metal particles without a shell [37]. However, the derivation provided here is more general. It considers the two incident field polarizations and by taking advantage of the effective medium representation of the particle polarizability can be used to describe interactions with particles consisting of unrestricted number of shells [35, 36]. The dipole-dipole model properly accounts for the SEF effect at sufficiently large separation distances in which case equation (3) simplifies as $F \to 1 + \frac{5}{2}|\mathcal{L}_p|^2 (a/R)^6$, emphasizing the importance of the nanoparticle dipolar response. However, as shown below, the dipole-dipole approximation fails when the emitter is in close proximity to the nanoparticle surface and non-radiating coupling (quenching) into higher order SP modes dominate the system response.

*2.2. Self-consistent multipole theory: quasi-static limit.*

To account for the non-radiative processes at short particle-emitter separation a general multi-mode theory must be developed. Here, we restrict our analyses to the quasi-static limit introducing the local electric potentials in the nanoparticle core, shell and free space

$$\phi_{core}(\vec{r}) = \sum_{l=0}^{\infty}\sum_{m=-l}^{l} A_{lm} r^l Y_l^m(\theta, \phi)$$
$$\phi_{shell}(\vec{r}) = \sum_{l=0}^{\infty}\sum_{m=-l}^{l} \left(B_{lm} r^l + \frac{C_{lm} a^{2l+1}}{r^{l+1}}\right) Y_l^m(\theta, \phi) \quad (4)$$
$$\phi_{out}(\vec{r}) = \phi_0(\vec{r}) + \phi_d(\vec{r}) + \sum_{l=0}^{\infty}\sum_{m=-l}^{l} \frac{D_{lm} a^{2l+1}}{r^{l+1}} Y_l^m(\theta, \phi)$$

The total potential outside the particle is split into three contributions, due to the external (laser) radiation $\phi_0(\vec{r})$, the molecular emitter $\phi_d(\vec{r})$, and nanoparticle. For $r < R$, the external and emitter potentials can be written in spherical harmonics (appendix A) as

$$\phi_0(\vec{r}) = \sum_{l=0}^{\infty} \sum_{m=-l}^{l} I_{lm}^0 r^l Y_l^m(\theta, \phi)$$

$$\phi_d(\vec{r}) = \sum_{l=0}^{\infty} \sum_{m=-l}^{l} I_{lm}^d r^l Y_l^m(\theta, \phi) \quad (r < R) \quad (5)$$

The multipole coefficients depend on the incident electric field $\vec{E}_0 = (E_{0,x}, E_{0,y}, E_{0,z})$, distance between emitter and particle surface and the emitter dipole moment $\vec{p}_d = (p_x, p_y, p_z)$ as

$$I_{lm}^0 = \delta_{l,1}\left(E_{0,x}\frac{\delta_{m,1} - \delta_{m,-1}}{2Y_{l,1}} + E_{0,y}\frac{\delta_{m,1} + \delta_{m,-1}}{2iY_{l,1}} - E_{0,z}\frac{\delta_{m,0}}{Y_{l,0}}\right)$$

$$I_{lm}^d = -\frac{1}{4\pi\epsilon_0 R^{l+2}}\left(p_x\frac{\delta_{m,1} - \delta_{m,-1}}{2Y_{l,1}} + p_y\frac{\delta_{m,1} + \delta_{m,-1}}{2iY_{l,1}} + p_z(l+1)\frac{\delta_{m,0}}{Y_{l,0}}\right), \quad (6)$$

where $Y_{l,0} = \sqrt{(2l+1)/4\pi}$ and $Y_{l,1} = Y_{l,0}/\sqrt{l(l+1)}$. Substituting the potentials equation (5) in equation (4) and enforcing the boundary conditions at the nanoparticle surfaces we obtain the remaining expansion coefficients as

$$A_{lm} = \frac{\epsilon_2(2l+1)(I_{lm}^0 + I_{lm}^d + D_{lm})}{\epsilon_2 + l(\epsilon_2 + \epsilon_1) + l(\epsilon_2 - \epsilon_1)p_l}$$

$$B_{lm} = \frac{(\epsilon_2 + l(\epsilon_2 + \epsilon_1))(I_{lm}^0 + I_{lm}^d + D_{lm})}{\epsilon_2 + l(\epsilon_2 + \epsilon_1) + l(\epsilon_2 - \epsilon_1)p_l}, \quad (7)$$

$$C_{lm} = p_l\frac{l(\epsilon_2 - \epsilon_1)(I_{lm}^0 + I_{lm}^d + D_{lm})}{\epsilon_2 + l(\epsilon_2 + \epsilon_1) + l(\epsilon_2 - \epsilon_1)p_l}$$

$$D_{lm} = -\mathcal{L}_{p,l}(I_{lm}^0 + I_{lm}^d)$$

where $p_l = p^{(2l+1)/3}$, $\mathcal{L}_{p,l} = l(\tilde{\epsilon}_l - \epsilon_0)/(l\tilde{\epsilon}_l + (l+1)\epsilon_0)$ is the polarizability line-shape function of the $l$-th multipole and we have introduced the mode specific effective permittivity [35]

$$\tilde{\epsilon}_l = \epsilon_2\frac{\epsilon_2 + l(\epsilon_2 + \epsilon_1) - (l+1)(\epsilon_2 - \epsilon_1)p_l}{\epsilon_2 + l(\epsilon_2 + \epsilon_1) + l(\epsilon_2 - \epsilon_1)p_l}. \quad (8)$$

Note that equation (8) coincides with equation (1) for the dipole response $\tilde{\epsilon}_1 = \epsilon_e$ ($l = 1$) as it should. To obtain self-consistently the induced dipole moment $\vec{p}_d$ (determining the $I_{lm}^d$ coefficients) we need to find the electric field $\vec{E}_d = -\vec{\nabla}(\phi_{out}(\vec{r}) - \phi_d(\vec{r}))\big|_{\vec{r}=\vec{R}}$ at the position of the dipole (excluding the emitter self-field) and relate it to the dipole moment using the linear response $\vec{p}_d = \alpha_d(\omega)\vec{E}_d$. A straightforward calculation is performed in appendix B and gives the relationships

$$p_x = \alpha_d E_{d,x} = \alpha_d \left( \frac{1 - \mu_p \mathcal{L}_p}{1 - \mu_p \mu_d \mathcal{L}_p \mathcal{L}_d s_\parallel} \right) E_{0,x}$$

$$p_y = \alpha_d E_{d,y} = \alpha_d \left( \frac{1 - \mu_p \mathcal{L}_p}{1 - \mu_p \mu_d \mathcal{L}_p \mathcal{L}_d s_\parallel} \right) E_{0,y} . \quad (9)$$

$$p_z = \alpha_d E_{d,z} = \alpha_d \left( \frac{1 + 2\mu_p \mathcal{L}_p}{1 - 4\mu_p \mu_d \mathcal{L}_p \mathcal{L}_d s_\perp} \right) E_{0,z}$$

where

$$s_\perp = \frac{1}{4} \sum_{l=1}^{\infty} \frac{\mathcal{L}_{p,l}(l+1)^2}{\mathcal{L}_p} \left(\frac{a}{R}\right)^{2l-2}$$

$$s_\parallel = \frac{1}{2} \sum_{l=1}^{\infty} \frac{\mathcal{L}_{p,l} l(l+1)}{\mathcal{L}_p} \left(\frac{a}{R}\right)^{2l-2} . \quad (10)$$

Using equation (9) and averaging over the incident field polarizations the fluorescence enhancement factor follows as

$$F = F_\parallel + F_\perp = \frac{2}{3} \left| \frac{1 - \mu_p \mathcal{L}_p}{1 - \mu_p \mu_d \mathcal{L}_p \mathcal{L}_d s_\parallel} \right|^2 + \frac{1}{3} \left| \frac{1 + 2\mu_p \mathcal{L}_p}{1 - 4\mu_p \mu_d \mathcal{L}_p \mathcal{L}_d s_\perp} \right|^2 , \quad (11)$$

where we have identified separately the contributions due to field components that are parallel ($F_\parallel$) and normal ($F_\perp$) to the nanoparticle surface. Since for $R \gg a$ the two sums given in equation (10) approach unity, in this limit the fluorescence enhancement factor asymptotically approaches the dipole-dipole result equation (3) as it should. For $R \ll a$, the factors $s_\parallel$ and $s_\perp$ are no longer unity and describe the influence of higher order SP multipoles on the fluorescence emission. It must be noted that equation (11) also holds for particles consisting of multiple shells provided the effective permittivities $\tilde{\epsilon}_l$ are modified according to [36]. We must note that to the best of our knowledge equation (11) provides the first exact self-consistent multi-mode treatment of molecular fluorescent enhancement due to core-shell structures in the quasi-static case.

## 3. Results and discussion

In this section the fluorescence enhancement is studies for nanoparticles immersed in a rhodamine B (RhB) dye solution. This particular dye has been investigated extensively in the literature and its emission properties are readily obtained. In what follows we implement the Lorentzian model due to [38], with $\omega_0 = 2.13$ eV and $\omega_{\tau,d} = 0.16$ eV. For the metal particle permittivity $\epsilon_m$ we use directly the experimental data due to Johnson and Christy [39]. In all simulations we restrict our studies to the case where the laser excitation frequency coincides with the emitter resonance frequency $\omega = \omega_0$, thus neglecting the Stokes shift which for RhB is small ($\Delta \omega < 90$ meV) . Under this condition the emitter line shape function is reduced to $\mathcal{L}_d = i\omega_0/\omega_{\tau,d}$ and in what follows for convenience we use the rendering $\mu_d \mathcal{L}_d = i(\tilde{a}_d/R)^3$, where $\tilde{a}_d = a_d(\omega_0/\omega_{\tau,d})^{1/3}$ is the reduced molecular dipolar radius.

*3.1. Single nanoparticle*

Here we study SEF due to a single metal nanoparticle without a shell. In this case we have $\tilde{\epsilon}_l = \epsilon_2 = \epsilon_1 = \epsilon_m$ and the sums in equation (10) can be represented explicitly through the incomplete beta function

$$s_\| = \frac{q\big((2\epsilon + 1 + s)B_s(q,-2) - \epsilon B_s(q,-1)\big)}{(\epsilon + 1)(1 - s)s^q}$$

$$s_\perp = \frac{s_\|}{2} + \frac{qB_s(q,-2)}{2s^q} \quad , \quad (12)$$

where $\epsilon = \epsilon_m/\epsilon_0$, $s = (a/R)^2$, and $q = (\epsilon + 2)/(\epsilon + 1)$. To demonstrate the theory, in figure 2 we consider the fluorescence enhancement of a single RhB molecule as function of the distance, $h = R - a$, to the nanoparticle surface. First we observe that the enhancement is rather weak having a maximal value of $F_{max} = 3.13$ for $h = 0.8$ nm. Such weak response is easy to explain by inspecting the emitter and particle dipolar polarizability line-shape functions (shown as an insert). For optimal enhancement the two line shapes must coincide. However, for the RhB molecule there is a considerable frequency mismatch between the two with the fluorescence radiation centered at $\omega_0 = 2.13$ eV while the metal particle response manifests a resonant behavior at the corresponding surface plasmon (SP) frequency $\omega_{sp} = 3.5$ eV. In figure 2 we have also included the dipole-dipole theory according equation (3). As expected, this approximate result coincides with the exact only at sufficiently large separation distances and does not capture the effect of radiation quenching into higher order non-radiative SP multipoles. To better understand the behavior at close proximity to the particle surface, we expand the sums equation (12) in terms of $h/a \ll 1$. In this limit the multipole correction factors can be simplified as

$$\lim_{h \to 0} s_\| = \left(\frac{\epsilon + 2}{\epsilon + 1}\right)\left(\frac{a}{2h}\right)^3$$

$$\lim_{h \to 0} s_\perp = \frac{s_\|}{2} \quad , \quad (13)$$

and the fluorescence enhancement has a simple asymptotic form

$$\lim_{h \to 0} F = \frac{3}{4}\left(\frac{2h}{\tilde{a}_d}\right)^6 \left|\frac{\epsilon + 1}{\epsilon - 1}\right|^2 \left(\frac{8 + |\epsilon|^2}{|\epsilon + 2|^2}\right). \quad (14)$$

The above result shows explicitly that the emission suppression (quenching) is proportional to the six power of the emitter-particle distance and thus is only present at vert short distances of the order of 1nm. Finally, it is important to note that the fluorescence enhancement is driven predominantly by the component of the electric field ($F_\perp$) that is perpendicular to the nanoparticle surface. This is expected since it is this field component that couples to the dominant dipolar SP mode.

*3.2. Nanoparticles with a single shell*

As discussed in the previous section optimal fluorescence enhancement can be achieved only if the metal nanoparticle SP resonance frequency coincides with the fluorescence frequency. This requirement is difficult to satisfy for metal nanoparticles without a shell even if we chose different metals. Hence, here we consider two single shell configurations; a metal-dielectric nanoparticle (MDNP) (the shell is dielectric) and a dielectric-metal nanoparticle (DMNP) (the shell is metallic). Setting the excitation frequency to coincide with that of the RhB emitter fixes the metal permittivity at $\epsilon_m(\omega = \omega_0) = -14.882 + 0.386i$ [39]. The effective permittivity of the system is then given by equation (1), with $\epsilon_1 = \epsilon_m$ and $\epsilon_2 = \epsilon_d$ for the MDNP and $\epsilon_2 = \epsilon_m$ and $\epsilon_1 = \epsilon_d$ for the DMNP. The nanoparticle dipolar polarizability line-shape functions $\mathcal{L}_p$ of the two shell configurations can readily be calculated and is shown in figure 3. There is a clear resonance behavior provided the SP resonance condition $\text{Re}(\epsilon_e(\omega_0, p)) = -2\epsilon_0$ is satisfied, which solved for the core filling fraction gives

$$p = \frac{1}{2}\text{Re}\left(\frac{(\epsilon_2 + 2\epsilon_0)(2\epsilon_2 + \epsilon_1)}{(\epsilon_2 - \epsilon_0)(\epsilon_2 - \epsilon_1)}\right), \quad (15)$$

Since $0 \leq p \leq 1$, this condition can be satisfied provided $-\epsilon_d < \frac{1}{2}\text{Re}(\epsilon_m(\omega_0)) < -\epsilon_0$ for the MDNP, and if $\text{Re}(\epsilon_m(\omega_0)) < \min\left(-\frac{1}{2}\epsilon_d, -2\epsilon_0\right)$ for the DMNP.

The SEF factor at the optimal core filling fraction equation (15) calculated as function of the distance between the molecule and particle surface is shown in figure 4. For the DMNP and MDNP configurations the sums in equation (10) no longer have solutions in simple functions and are obtained numerically. For each configuration we also consider three dielectric cases. Generally the DMNP configuration is found to provide higher fluorescence enhancement for dielectric cores having practically achievable permittivities. A general trend of increasing fluorescence is observed with decrease of the core dielectric constant. The opposite behavior is found for the MDNP configurations, which is in agreement with the behavior of the nanoparticle dipolar line-shape function studied in figure 3. In both cases the fluorescence enhancement is substantially higher compared to the single metal particle studied in the previous section.

Finally, we complete the discussion by considering a general emitter with varying emission frequency $\omega_0$ equal to that of the excitation laser radiation. The results are presented in figure 5 with all calculations performed at the corresponding optimal core filling fractions and separation distances. We observe a strong frequency dependence of the maximal SEF. The DMNP configuration presents stronger effect for low and moderate frequencies $\omega_0 < 2\text{eV}$, while the MDNP is applicable only within a narrow frequency range $2\text{eV} < \omega_0 < 3.5\text{eV}$ in accordance with the applicability condition of equation (15). The artificial cut-off for $\omega_0 < 0.64\text{eV}$ is due to the absence of experimental data for the silver permittivity which is sourced from Johnson and Christy [39]. It must be noted that the observed decrease of the SEF factors for $\omega_0 > 2\text{eV}$ is due to increased losses associated with interband transitions in the metal. Within this region both configurations provide similar SEF factors. Overall, the two core-shell configurations over perform the single metal nanoparticle case for all emitter frequencies. While the optimal core filling fraction, see figure 5(b) and figure 5(d), spans the entire possible range $0 < p_{opt} < 1$, the optimal emitter-particle separation distance $h_{opt}$ is found to be weakly dependent on the emission frequency and dielectric permittivity and is or the order of 1nm. This can be verified by considering the short distance asymptotic equation (13), which under the substitution $\epsilon \to \epsilon_2/\epsilon_0$ holds even for the core-shell nanoparticles. Taking advantage of the small parameter $h/a \ll 1$ and working with equation (11) we obtain an approximated result for the optimal separation distance $h_{opt} \approx \tilde{a}_d|(\epsilon_2 - \epsilon_0)/(\epsilon_2 + \epsilon_0)|^{1/3} \approx \tilde{a}_d$. The maximal SEF can also be estimated and is found to follow closely the particle line shape-function at resonance $F_{max} \approx \frac{3}{2}|\mathcal{L}_p(\omega_0, p_{opt})|^2$. Finally, in figures 5(c)-(f) we compare the SEF factor calculated self-consistently and based on the traditional non self-consistent approach ($s_\perp = s_\parallel = 0$). For all cases and parameters used the self-consistent SEF factor is found to be weaker. This is expected due to the self-coupling of the dipolar radiation which tends to decrease the effective dipole moment according to equation (9).

We would like to complete the discussion by considering possible shortcoming of the provided model. As stated in the beginning of the paper, the model is based on the quasi-static approximation and the theory can only be applied for particle-emitter separation distances $h < c/\omega$. For the rhodamine B (RhB) molecule this restriction translates to $h < 92$ nm, with the presented results (see Figures (2)-(4)) being well within the applicability of the model. Furthermore, at higher separation distances where retardation effects start to dominate, the SEF factor is expected to be small in general due to the $1/h^2$ dependence of the power radiated by the dipole that is coupled to the particle. However, it may still be worth considering the extension of the current work to include retardation. This can be achieved based on an extension of the quasi-effective medium theory [36] for 3D multi-layer magneto-dielectric systems, with the effective permittivity equation (8) now written as function of the wavevector. Additionally, when the metal particle size or shell thickness become of the order, $2a \sim v_F/\omega$, where $v_F$ is the Fermi velocity, non-local elects will also be expected [40-43]. For silver and the particle size $2a = 30 nm$ considered in this paper, we have $2a\omega/v_F \gg 1$ and nonlocal effects are expected to be weak. However, for smaller particles and thinner shells with $p < (v_F/2a\omega)^3$ (for MDNP) and $p > (1 - v_F/2a\omega)^3$ (for DMNP) non-local effects must be taken into account.

## 4. Conclusions

In this paper, we present a self-constant analytical theory of molecular fluorescence enhancement due to metal-dielectric nanoparticles. The theory provides explicit insights related to the surface plasmon (SP) mechanism behind the phenomenon. Specifically, it shows that emitters that are sufficiently far from the nanoparticle can be described by a simple dipole-dipole model while at short distances the fluorescence emission is quenched as result of coupling to higher-order non-radiative SP multipoles. The general condition for matching the emitter and nanoparticle line shape functions is derived. The core-shell configurations that provide optimal surface enhanced fluorescence (SEF) are identified and should be considered for applications related to fluorescence spectroscopy and bio-sensing.

**Data availability statement**

The data that support the findings of this study are available upon reasonable request from the authors.

**Acknowledgments**

**Appendix A: Potentials and spherical harmonics**

To obtain a close form solution of equation (4) it is convenient to represent the incident and dipole potentials in spherical harmonics. For the incident potential it is straightforward to show by inspection that

$$\phi_0(\vec{r}) = -\vec{r} \cdot \vec{E}_0 = -r\sin(\theta)\left(E_{0,x}\cos(\phi) + E_{0,y}\sin(\phi)\right) - r\cos(\theta)E_{0,z}$$

$$= \sum_{l=0}^{\infty}\sum_{m=-l}^{l} I_{lm}^0 r^l Y_l^m(\theta,\phi), \quad (A.1)$$

where the explicit form of the expansion coefficients $I_{lm}^0$ is given by equation (6). The dipole potential of the emitter has the traditional form

$$\phi_d(\vec{r}) = \frac{\vec{p}_d \cdot (\vec{r}-\vec{R})}{4\pi\epsilon_0|\vec{r}-\vec{R}|^3} = \frac{\left(p_x\cos(\phi)+p_y\sin(\phi)\right)r\sin(\theta) + p_z(r\cos(\theta)-R)}{4\pi\epsilon_0(r^2 - 2rR\cos(\theta) + R^2)^{3/2}}, \quad (A.2)$$

where $\vec{p}_d = (p_x, p_y, p_z)$ is the dipole moment and $\vec{R} = R\hat{z}$. For $r < R$ we implement the series expansion

$$\frac{1}{\sqrt{r^2+R^2-2rR\cos(\theta)}} = \sum_{l=0}^{\infty} \frac{r^l}{R^{l+1}} P_l^0(\cos(\theta)), \quad (A.3)$$

and write

$$\frac{r\cos(\theta)-R}{(r^2-2rR\cos(\theta)+R^2)^{3/2}} = \frac{\partial}{\partial R}\left(\frac{1}{\sqrt{r^2+R^2-2rR\cos(\theta)}}\right) = -\sum_{l=0}^{\infty}\frac{(l+1)r^l}{R^{l+2}}P_l^0(\cos(\theta))$$

$$\frac{r\sin(\theta)}{(r^2-2rR\cos(\theta)+R^2)^{3/2}} = -\frac{1}{R}\frac{\partial}{\partial\theta}\left(\frac{1}{\sqrt{r^2+R^2-2rR\cos(\theta)}}\right) = -\sum_{l=0}^{\infty}\frac{r^l}{R^{l+2}}P_l^1(\cos(\theta))$$

$$. \quad (A.4)$$

Substituting equation (A.4) in equation (A.2) the dipole potential follows as

$$\phi_d(\vec{r}) = -\frac{1}{4\pi\epsilon_0}\sum_{l=0}^{\infty}\frac{r^l}{R^{l+2}}\left(\left(p_x\cos(\phi)+p_y\sin(\phi)\right)P_l^1(\cos(\theta)) + p_z(l+1)P_l^0(\cos(\theta))\right), \quad (A.5)$$

Recasting equation (A.5) in spherical harmonics we arrive at

$$\phi_d(\vec{r}) = \sum_{l=0}^{\infty}\sum_{m=-l}^{l} I_{lm}^d r^l Y_l^m(\theta,\phi), \quad (A.6)$$

## Appendix B: Self-consistent field equations

The electric field at the position of the molecular emitter (excluding the emitter self-field) is obtained from the corresponding potential

$$\tilde{\phi}(\vec{r}) = \phi_{out}(\vec{r}) - \phi_d(\vec{r}) = \phi_0(\vec{r}) + \sum_{l=0}^{\infty}\sum_{m=-l}^{l} \frac{D_{lm} a^{2l+1}}{r^{l+1}} Y_l^m(\theta, \phi). \quad (B.1)$$

Substituting the expansion coefficients equation (6) we obtain

$$\tilde{\phi}(\vec{r}) = \phi_0(\vec{r}) - \sum_{l=1}^{\infty} \frac{\mathcal{L}_{p,l} a^{2l+1}}{r^{l+1}} \Big( \big(a_{l,x}\cos(\phi) + a_{l,y}\sin(\phi)\big) P_l^1(\cos(\theta))$$
$$- (l+1) a_{l,z} P_l^0(\cos(\theta)) \Big), \quad (B.2)$$

where

$$a_{l,x} = \delta_{l,1} E_{0,x} - \frac{p_x}{4\pi\epsilon_0 R^{l+2}}$$
$$a_{l,y} = \delta_{l,1} E_{0,y} - \frac{p_y}{4\pi\epsilon_0 R^{l+2}}. \quad (B.3)$$
$$a_{l,z} = \frac{1}{2}\delta_{l,1} E_{0,z} + \frac{p_z}{4\pi\epsilon_0 R^{l+2}}$$

The electric field then follows $\vec{E} = -\vec{\nabla}\tilde{\phi} = \vec{E}_0 + E_r \hat{r} + E_\theta \hat{\theta} + E_\phi \hat{\phi}$, with the field components given as

$$E_r = \sum_{l=1}^{\infty} \frac{(l+1)\mathcal{L}_{p,l} a^{2l+1}}{r^{l+2}} \Big( (l+1) a_{l,z} P_l^0(\cos(\theta)) - \big(a_{l,x}\cos(\phi) + a_{l,y}\sin(\phi)\big) P_l^1(\cos(\theta)) \Big)$$
$$E_\theta = \sum_{l=1}^{\infty} \frac{\mathcal{L}_{p,l} a^{2l+1}}{r^{l+2}} \left( \big(a_{l,x}\cos(\phi) + a_{l,y}\sin(\phi)\big) \frac{\partial P_l^1(\cos(\theta))}{\partial\theta} - (l+1) a_{l,z} \frac{\partial P_l^0(\cos(\theta))}{\partial\theta} \right). \quad (B.4)$$
$$E_\phi = \sum_{l=1}^{\infty} \frac{\mathcal{L}_{p,l} a^{2l+1}}{r^{l+2}} \big(a_{l,y}\cos(\phi) - a_{l,x}\sin(\phi)\big) \frac{P_l^1(\cos(\theta))}{\sin(\theta)}$$

Using the asymptotic forms of the associated Legendre polynomials

$$\lim_{\theta\to 0} P_l^0(\cos(\theta)) = 1, \quad \lim_{\theta\to 0} P_l^1(\cos(\theta)) = \lim_{\theta\to 0} \frac{\partial P_l^0(\cos(\theta))}{\partial\theta} = 0$$
$$\lim_{\theta\to 0} \frac{P_l^1(\cos(\theta))}{\sin(\theta)} = \lim_{\theta\to 0} \frac{\partial P_l^1(\cos(\theta))}{\partial\theta} = -\frac{l(l+1)}{2} \quad, \quad (B.5)$$

we can recast the electric field components equation (B.4) as

$$E_r = \sum_{l=1}^{\infty} \frac{(l+1)^2 \mathcal{L}_{p,l} a^{2l+1}}{R^{l+2}} a_{l,z}$$

$$E_\theta = -\sum_{l=1}^{\infty} \frac{l(l+1)\mathcal{L}_{p,l} a^{2l+1}}{2R^{l+2}} \left(a_{l,x}\cos(\phi) + a_{l,y}\sin(\phi)\right). \quad (B.6)$$

$$E_\phi = \sum_{l=1}^{\infty} \frac{l(l+1)\mathcal{L}_{p,l} a^{2l+1}}{2R^{l+2}} \left(a_{l,x}\sin(\phi) - a_{l,y}\cos(\phi)\right)$$

Reversing to Cartesian coordinates we obtain

$$\vec{E}_d = \vec{E}_0 + E_r \hat{z} + \left(E_\theta \cos(\phi) - E_\phi \sin(\phi)\right)\hat{x} + \left(E_\theta \sin(\phi) + E_\phi \cos(\phi)\right)\hat{y}$$

$$= \vec{E}_0 - \sum_{l=1}^{\infty} \frac{l(l+1)\mathcal{L}_{p,l} a^{2l+1}}{R^{l+2}} \left(\frac{a_{l,x}\hat{x} + a_{l,y}\hat{y}}{2} - \frac{l+1}{l} a_{l,z}\hat{z}\right). \quad (B.7)$$

Substituting equation (B.3) in equation (B.7) and rearranging we end up with the governing equation of the emitter's self-consistent dipole moment

$$\vec{p}_d = \alpha_d \vec{E}_d = \alpha_d \left((1 - \mu_p \mathcal{L}_p)E_{0,x} + \frac{\mu_p \mathcal{L}_p s_\parallel}{4\pi\epsilon_0 R^3} p_x\right)\hat{x} + \alpha_d \left((1 - \mu_p \mathcal{L}_p)E_{0,y} + \frac{\mu_p \mathcal{L}_p s_\parallel}{4\pi\epsilon_0 R^3} p_y\right)\hat{y}$$

$$+ \alpha_d \left((1 + 2\mu_p \mathcal{L}_p)E_{0,z} + \frac{\mu_p \mathcal{L}_p s_\perp}{\pi\epsilon_0 R^3} p_z\right)\hat{z}. \quad (B.8)$$

where $\mu_p = (a/R)^3$, and the modal functions $s_\parallel$ and $s_\perp$ are given by equation (10). Form equation (B.9) the components of the self-consistent dipole moment are easily obtained and are given by equation (9).

**References**

[1] Shahzad A, Edetsberger M, and Koehler G 2010 Appl. Spectrosc. Rev. **45** 1

[2] Haustein E, and Schwille P 2003 Methods **29** 153

[4] Ha T and Tinnefeld P 2012 Annu. Rev. Phys. Chem. **63** 595

[5] Sultangaziyev A and Bukasov R 2020 Sens. Bio-Sens. Res. **30** 100382

[6] Shank C V 1975 Rev. Mod. Phys. **47** 649

[7]  Bimberg D and Pohl U W 2011 Mater. Today **14** 388 (2011)

[8] Cortes E, Etchegoin P G, Le Ru E C, Fainstein A, Vela M E, and Salvarezza R C 2010 J. Am. Chem. Soc. **132** 18034

[9] Purcell E M 1964 Phys. Rev. **69** 681

[10] Genov D A, Oulton R F, Bartal G and Zhang X 2011 Phys. Rev. B **83** 245312

[11] Liebermann T and Knoll W 2000 Colloids Surf. A Physicochem. Eng. Asp. **171** 115

[12] Kano H and Kawate S 1996 Opt. Lett. **21** 1848

[13] Kneipp K, Wang Y, Kneipp H, Perelman L T, Itzkan I, Dasari R R and Feld M S 1997 Phys. Rev. Lett. **78** 1667

[14] Genov D A, Sarychev A K, Shalaev V M and Wei A 2004 Nano Lett. **4** 153

[15] Bergman D J and Stockman M I 2003 Phys. Rev. Lett. **90** 027402

[16] Okamoto K, Niki I, Shvartser A, Narukawa Y, Mikai T and Scherer A 2004 Nature Mater. **3** 601.

[17] Gu Q, Slutsky B, Vallini F, Smalley J S T, Nezhad M P, Frateschi N C and Fainman Y 2013 Opt. Express **21** 15604

[18] Ambati M, Genov D A, Oulton R and Zhang X 2008 IEEE J. Sel. Top. Quantum Electron **14** 1395

[19] Azzam S I, Kildishev A V, Ma R M, Ning C Z, Oulton R, Shalaev V M, Stockman M I, Xu J L and Zhang X 2020 Light Sci. Appl. **9** 90

[20] Chance R R, Prock A and Silbey R 1975 J. Chem. Phys. **62** 2245

[21] Chance R R, Prock A and Silbey R 1975 Phys. Rev. A **12** 1448

[22] Gersten J and Nitzan A 1981 J. Phys. Chem. **75** 1139

[23] Ruppin R 1982 J. Chem. Phys. **76** 1681

[24] Chew H 1987 J. Chem. Phys. **87** 1355

[25] Kim Y S, Leung P and George T F 1988 Surf. Sci. **195** 1

[26] Ringler M, Schwemer A, Wunderlich M, Nichtl A, Kurzinger K, Klar T A and Feldmann J 2008 Phys. Rev. Lett. **100** 203002

[27] Le K Q 2015 Plasmonics 10, 475

[28] Chew H, Kerker M and McNulty P J 1976 J. Opt. Soc. Am. **66** 440

[29] Lavallard P, Rosenbauer M, and Gacoin T 1996 Phys. Rev. A **54** 5450

[30] Moroz A 2005 Chem. Phys. **317** 1

[31] Agarwal G S 1975 Phys. Rev. A **12** 1475-1497

[32] Wylie J M and Sipe J E 1984 Phys. Rev. A **30** 1185

[33] Johansson P, Xu H and Kall M 2005 Phys. Rev. B **72** 035407

[34] Wei Y, Li L, Sun D X, Wang M L, and Zhu Y Y 2018 Sci. Rep. **8** 1832

[35] Mundru P C, Pappakrishnan V K and Genov D A 2012 Phys. Rev. B **85** 045402

[36] Genov D A and Mundru P C 2014 J. Opt. **16** 015101

[37] Simovski C 2015 Photonics **2** 568

[38] Craighead H G and Glass A M 1981 Opt. Lett. **6** 248

[39] Johnson P B and Christy E W 1972 Phys. Rev. B **6** 4370

[40] Abajo F J G D 2008 J. Phys. Chem. C **112** 17983

[41] David C and Abajo F J G D 2011 J. Phys. Chem. C **115** 19470

[42] Raza S, Bozhevolnyi S I, Wubs M and Mortensen N A 2015 J. Phys. Condens. Matter **27** 183204

[43] Ching H Y, Guo G Y, Chiang H P, Tsai D P and Leung P T 2010 Phys. Rev. B **82** 165440

**Figures**

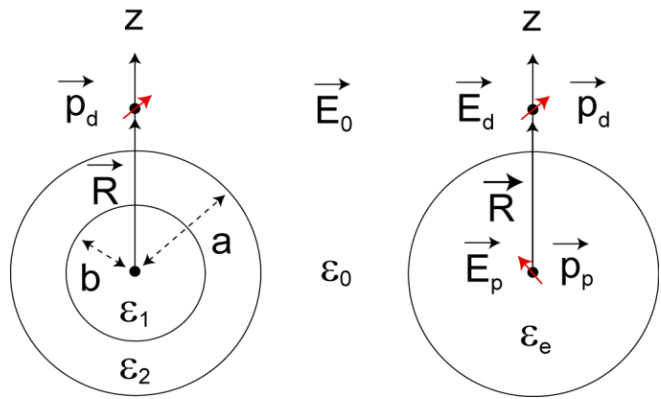

**Figure 1.** (left) Two-dimensional schematic of a dielectric core-shell nanoparticle in the presence of a point dipole. (right) Effective medium dipole-dipole model of the same configuration.

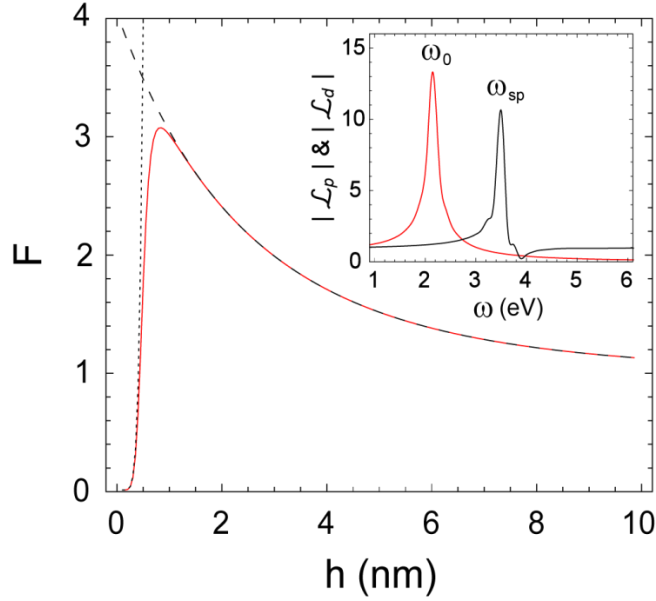

**Figure 2.** Self-consistent fluorescence enhancement factor calculated at the emitter frequency $\omega_0 = 2.13 \text{ eV}$ of a rhodamine B (RhB) molecule as function of the distance to a silver nanoparticle with radius $a = 15 nm$. The large distance asymptotic (dipole-dipole) result given by equation (3) is depicted with dashed line while the short-distance asymptotic equation (14) is represented with dotted line. The line-shape functions of the emitter (red line) and nanoparticle (black line) are shown in the insert.

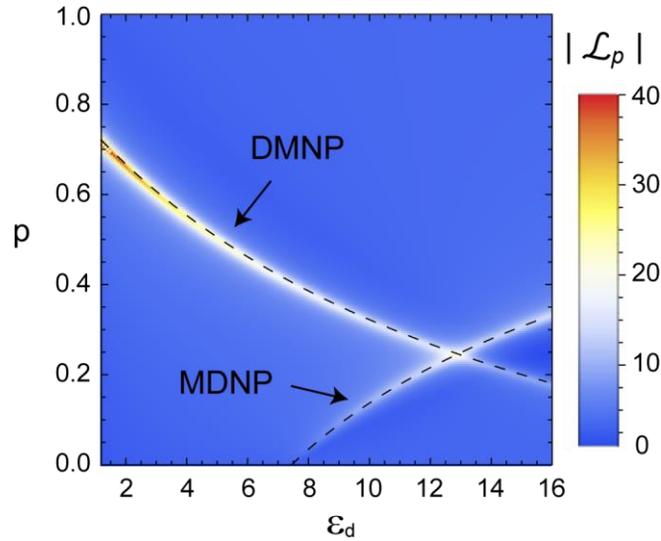

**Figure 3.** Dipolar polarizability line-shape functions at the emitter frequency $\omega_0 = 2.13 \text{ eV}$ of a rhodamine B (RhB) molecule calculated for silver based DMNP and MDNP configurations. The optimal filling fractions calculated using equation (15) are depicted with dashed black lines.

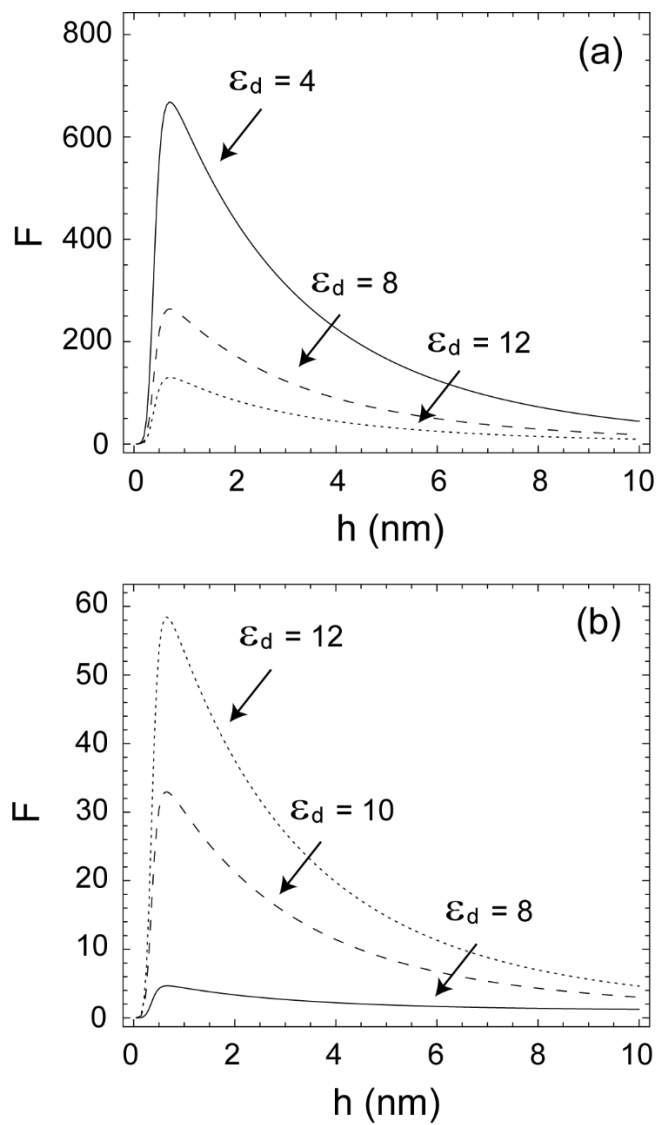

**Figure 4.** Self-consistent fluorescence enhancement factor calculated at the emitter frequency $\omega_0 = 2.13$ eV of a rhodamine B (RhB) molecule in proximity of silver based (a) DMNP and (b) MDNP configurations. The core filling fractions for the various dielectric permitivities are calculated using equation (15), and the nanoparticles radius is set at $a = 15 nm$.

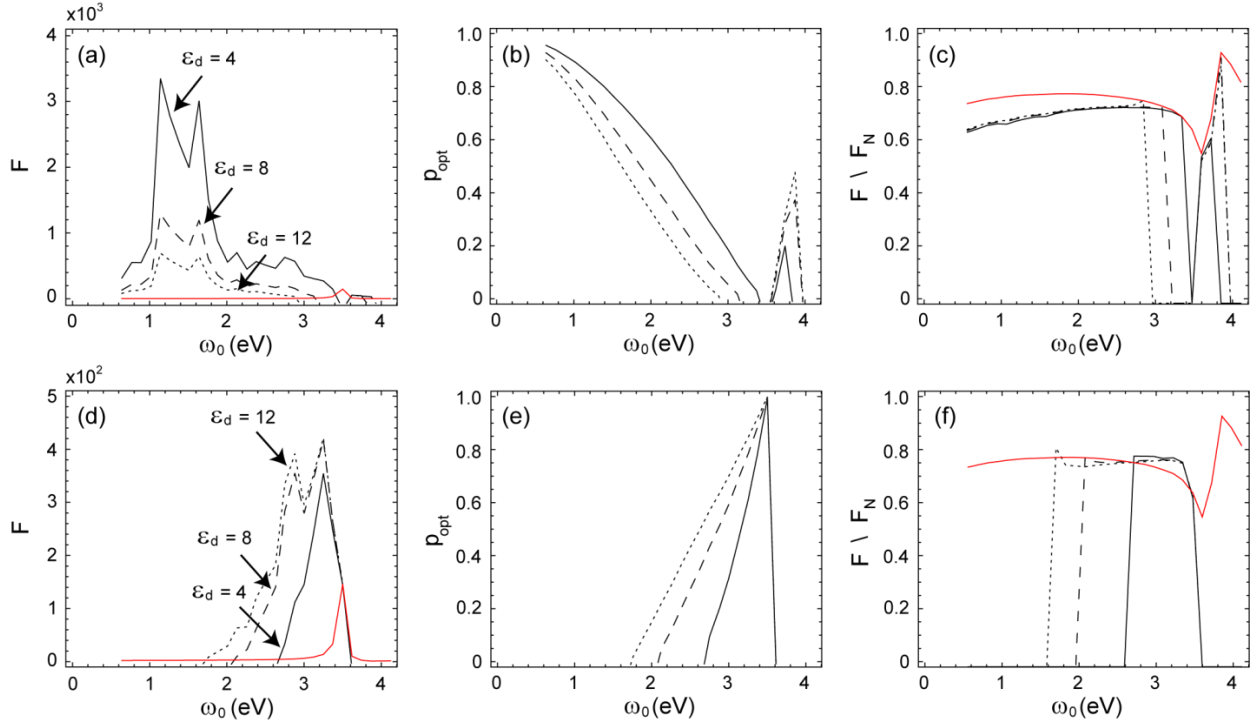

**Figure 5.** Maximal self-consistent fluorescence factor $F$, optimal core filling fraction $p_{opt}$, and ratio between $F$ and the non- self-consistent SEF factor $F_N$ calculated for the DMNP (a)-(c) and MDNP (d)-(f) configurations. In the calculations we vary the emitter frequency $\omega_0$ and consider three separate dielectric permittivities (solid, dashed and doted lines) with the single silver nanoparticle case included as a reference (solid red line). For all cases the emitter relaxation rate is fixed at $\omega_{\tau,d} = 0.16$ eV and the nanoparticles radius is set at $a = 15 nm$.